\documentclass{article}

\usepackage{arxiv}

\usepackage[utf8]{inputenc} 
\usepackage[T1]{fontenc}    
\usepackage{hyperref}       
\usepackage{url}            
\usepackage{booktabs}       
\usepackage{amsfonts}       
\usepackage{nicefrac}       
\usepackage{microtype}      
\usepackage{lipsum}		
\usepackage{graphicx}
\usepackage{doi}
\usepackage[sort&compress,super,numbers]{natbib}
\usepackage{amsmath}
\usepackage{multirow}
\usepackage[labelfont=bf]{caption}

\title{Net electrophilicity as computational route for the choice of favorable ionic liquids in nanoparticle production}

\author{ \href{https://orcid.org/my-orcid?orcid=0000-0002-9930-7728}{\includegraphics[scale=0.06]{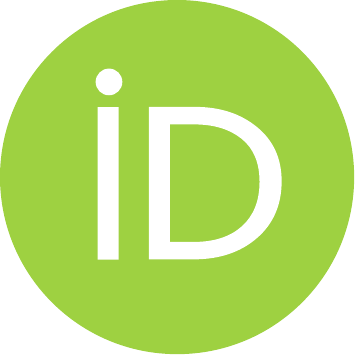}\hspace{1mm}Marta Bon} \\
	Electron Microscopy Center \\
    Empa, Swiss Federal Laboratories for Materials Science and Technology
    CH-8600 Dübendorf, Switzerland\\
    Exscientia\thanks{Current address} \\
    The Schrödinger Building\\
    Oxford Science Park\\ 
    Oxford OX4 4GE, United Kingdom
	\And
	\href{https://orcid.org/0000-0003-2343-7275}{\includegraphics[scale=0.06]{orcid.pdf}\hspace{1mm}Debora Keller} \\
    Electron Microscopy Center \\
	Empa, Swiss Federal Laboratories for Materials Science and Technology
    CH-8600 Dübendorf, Switzerland
    \And
    \href{https://orcid.org/0000-0003-2391-5943}{\includegraphics[scale=0.06]{orcid.pdf}\hspace{1mm}Rolf Erni} \\ 
	Electron Microscopy Center \\
    Empa, Swiss Federal Laboratories for Materials Science and Technology
    CH-8600 Dübendorf, Switzerland\\
	\And
	\href{https://orcid.org/0000-0001-5690-9000}{\includegraphics[scale=0.06]{orcid.pdf}\hspace{1mm}Daniele Passerone}\thanks{Corresponding author} \\ 
	nanotech@surfaces Laboratory\\
	Empa, Swiss Federal Laboratories for Materials Science and Technology
    CH-8600 Dübendorf, Switzerland\\
	\texttt{daniele.passerone@empa.ch}
}

\renewcommand{\shorttitle}{\textit{arXiv} Template}

\hypersetup{
pdftitle={A template for the arxiv style},
pdfsubject={q-bio.NC, q-bio.QM},
pdfauthor={David S.~Hippocampus, Elias D.~Striatum},
pdfkeywords={First keyword, Second keyword, More},
}

\begin{document}
\maketitle

\begin{abstract}
	In the last years, the potential of using ionic liquids (IL)s as an environment for nanoparticle (NP) synthesis has been demonstrated and in particular, triggering NP formation in ILs by electron irradiation has been reported as a very simple and clean route for NP production. Starting from the recent evidence for a correlation between an IL’s capability to support NP production and the radiochemical instability of the IL’s cation, we used conceptual Density Functional Theory (DFT) to provide a pre-screening of a set of different IL cations. The screened quantity is the net electrophilicity which we suggest as possible measure of this instability. Therefore, our work not only gives a measure for the likelihood of NP generation in different ILs, but it also provides a model which can further be extended and applied to obtain information about any other IL of interest. Moreover, our theoretical approach outlines a strategy which may reduce a lengthy experimental investigation for the identification of the most suitable IL for a particular reaction.
\end{abstract}


\begin{figure}[htp]
\centering
\includegraphics{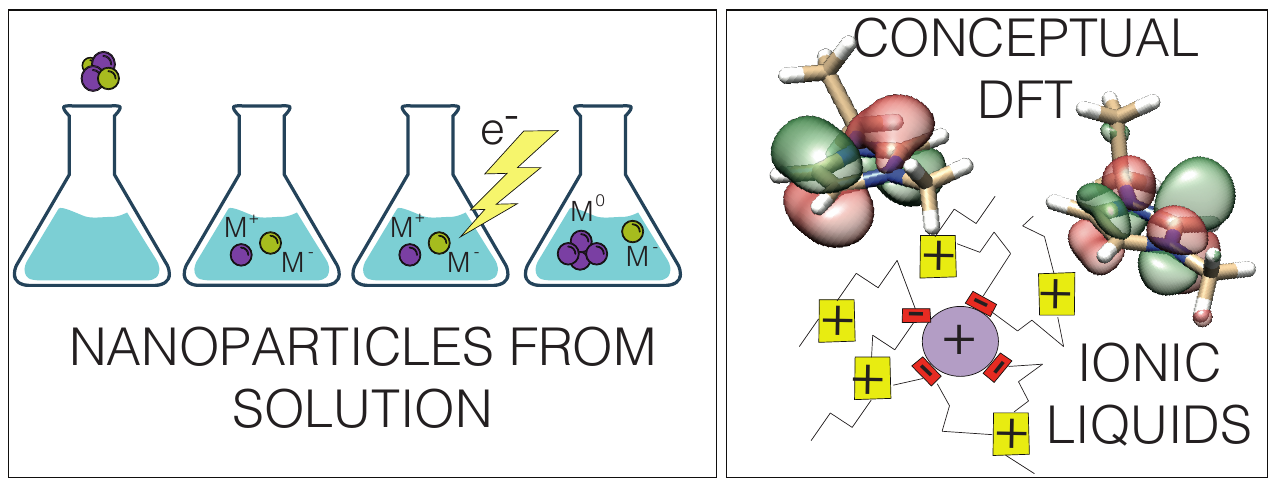} 
\caption{Graphical Table of Contents}
\end{figure}

\section{Introduction}

\begin{figure}[htp]
\centering
\includegraphics[width=0.5\columnwidth]{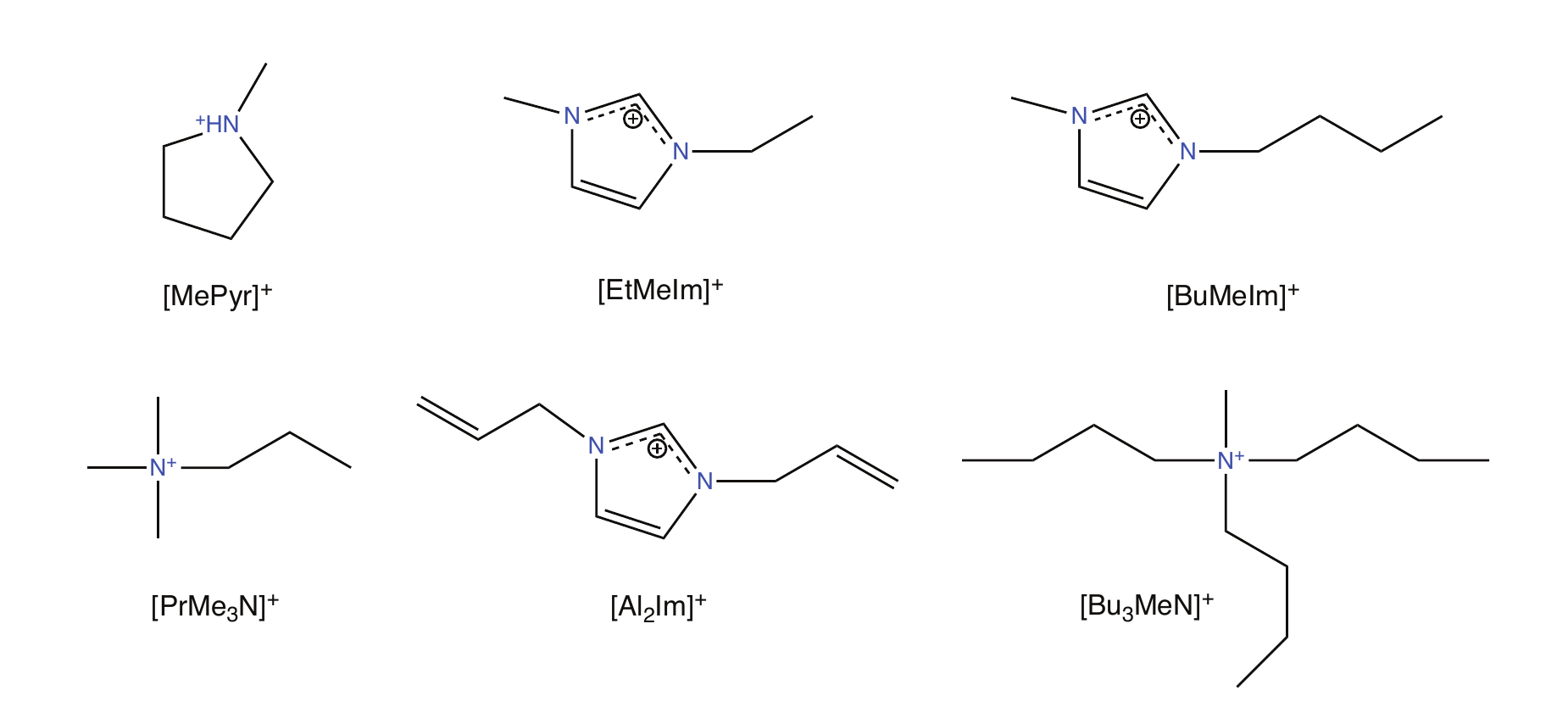} 
\caption{Cations constituting the ILs used for NP production triggered by irradiation (cfr. Table \ref{tab:ILsforNP}). The corresponding $\Delta \omega$ are reported in Figure \ref{fig:ILsBenchmark}.}
\label{fig: cations}
\end{figure}

In the last decades, nanotechnology has been spanning several industrial sectors, from energy storage to drug delivery.\cite{schmid2005nanoparticles} The wide range of possible applications together with their tunable properties make metal nanoparticles (NP) key materials for further progress in science and technology. Among the several bottom-up strategies for NP production, liquid phase processes have recently exploited the use of room-temperature ionic liquids (ILs) that allow the synthesis of pure NP, not converted with stabilizing agents that can negatively affect their catalytic activity. \cite{scheeren2006synthesis, tsuda2009gold, wishart2010ionic, khare2010strong, kuwabata2010room, tsuda2012various, uematsu2014atomic, kimura2014situ, patil2014shape, minamimoto2014fine, guleria2017probing} One possible way to trigger NP formation is to irradiate the IL, which contains the NP precursor, e.g. with electrons.\cite{kuwabata2010room} Regarding practical aspects, this process provides a simple and clean route for NP synthesis. More in general, this approach opens new possibilities in the field of NPs research. As ILs have the unique property to withstand high-vacuum conditions in their liquid state, the early stages of NP formation triggered by an electron beam can e.g. be studied by in situ transmission electron microscopy (TEM) at very high spatial and moderate temporal resolution.\cite{uematsu2014atomic} In a simplified way, the NP production in the IL can be described by a cascade of four subsequent steps: At first, the metal precursor (usually a salt) dissolves in the IL as metal cations and other ions (step 1). Second, the electron beam interacts with the IL, mainly producing solvated electrons and radical IL cations (step 2). Then, these highly reactive species interact with the metal cations and reduce them to neutral atoms (step 3). The metal atoms finally undergo a nucleation-and-growth reaction resulting in the formation of NPs (step 4). \\
In this respect, it has been recently underlined that while the shape of the NP is affected by the IL anion,\cite{khare2010strong,patil2014shape} the success of metal ion reduction and NP production mainly depends on the characteristics of the IL’s cation, and in particular on its radiochemical instability.\cite{tsuda2009gold, kuwabata2010room, uematsu2014atomic} In fact, as the solvated electrons and the IL’s radicals are expected to be responsible for reducing the metal precursor, their properties are likely to be critical for the NP formation mechanism. \\
Conceptual Density Functional Theory\cite{geerlings2003conceptual} qualifies as possible instrument to identify and rank the different cations on the basis of their reactivity, estimating the so called net electrophilicity index ($\Delta \omega$).\cite{chattaraj2009net}
This is a global reactivity indicator (one value is assigned for each system, i.e. molecule), and is defined as the difference between the electroaccepting ($\omega^+$) and the electrodonating ($\omega^-$) power, which are function of the ionization potentials (IP) and the electron affinity (EA):
\begin{equation}
\label{eq:deltaomega}
\Delta \omega=\omega^+  -(-\omega^-)\approx\frac{(IP+3EA)^2+(3IP+EA)^2}{16(IP-EA)}.
\end{equation}
High values of $\Delta \omega$ correspond to highly reactive species, low values to stable and inert molecules. The quantities in Eq. \ref{eq:deltaomega} are computed via DFT calculation (in this work, hybrid and double hybrid functionals are used, see Supporting Information (SI) for details) , and the word conceptual in the method stems from the fact that we extract a measurable quantity and relate it to a known chemical concept (e.g. chemical potential, chemical hardness, electrophilicity). Note that different definitions for $\omega^+$ and $\omega^-$ are provided in literature,\cite{gazquez2007electrodonating} and further discussion is reported in Sections \ref{section: theory} and \ref{section: test definitions} of the SI.\\
In this work, we rank the cationic species on the basis of $\Delta \omega$, and discuss the results in light of the IL power in NP production as found in published experimental work (Figures \ref{fig: cations} and \ref{fig:ILsBenchmark} and Table \ref{tab:ILsforNP}). We further extend our study to other IL cations that can be used in this field, offering the first pre-screening predictive study for this purpose (Figure \ref{fig:PreScreening}).

\begin{table}[htp]
\centering
\caption{List of ILs experimentally used for NP production. Structures are reported in the Figure \ref{fig: cations}.}
\label{tab:ILsforNP}
    \begin{tabular}{llll}
    \toprule
    IL Cation  & Anion  & Precursor  & NP  \\
    \midrule
    \multicolumn{4}{c}{IL effective in NP production} \\
    \midrule
    $\mathrm{[Al\_2Im]^+}$ & $\mathrm{[Br]^-}$ & $\mathrm{NaClO_3}$ & Na [\citenum{kimura2014situ}] \\
    \midrule
    \multirow{3}[0]{*}{$\mathrm{[Bu Me Im]^+}$} & \multirow{2}[0]{*}{$\mathrm{[NTf_2]^-}$} & $\mathrm{MgCl_2}$,$\mathrm{Fe[Tf_2N]_2}$,$\mathrm{Zn[Tf_2N]_2}$ & Pd, Pt [\citenum{tsuda2012various}]\\
          &       & $\mathrm{NaAuCl_4}$ & Au [\citenum{tsuda2009gold}] \\
          & $\mathrm{[PF6]^-}$  & $\mathrm{Pt_2dba)_3}$ & Pt[\citenum{scheeren2006synthesis}] \\
    \midrule
    \multirow{2}[0]{*}{$\mathrm{Et Me Im}^+$} & $\mathrm{NTf_2^-}$ & $\mathrm{K_2PdCl_4}$, $\mathrm{K_2PtCl_4}$  & Pd, Pt [\citenum{tsuda2012various}] \\
          & $\mathrm{Et SO_4^-}$ & $\mathrm{Na_2SeO_3}$ & Se [\citenum{guleria2017probing}] \\
    \midrule
    $\mathrm{MePyr^+}$ & $\mathrm{Ac^-}$ & $\mathrm{NO_3}$   & Ag [\citenum{patil2014shape}]  \\
    \midrule
    $\mathrm{PrMe_3N^+}$ & $\mathrm{NTf_2^-}$ & $\mathrm{Ag_2O}$  & Ag [\citenum{minamimoto2014fine}] \\
    \midrule
    \multicolumn{4}{c}{IL ineffective in NP production} \\
    \midrule
    $\mathrm{Bu Me N^+}$ & $\mathrm{NTf_2^-}$ & $\mathrm{K_2PdCl_4}$, $\mathrm{K_2PtCl_4}$  & Pd, Pt [\citenum{tsuda2012various}]\\
    \bottomrule
    \end{tabular}%
\end{table}

\section{Results}
\subsection{Benchmarking}
\begin{figure}[htp]
\centering
\includegraphics[width=0.5\columnwidth]{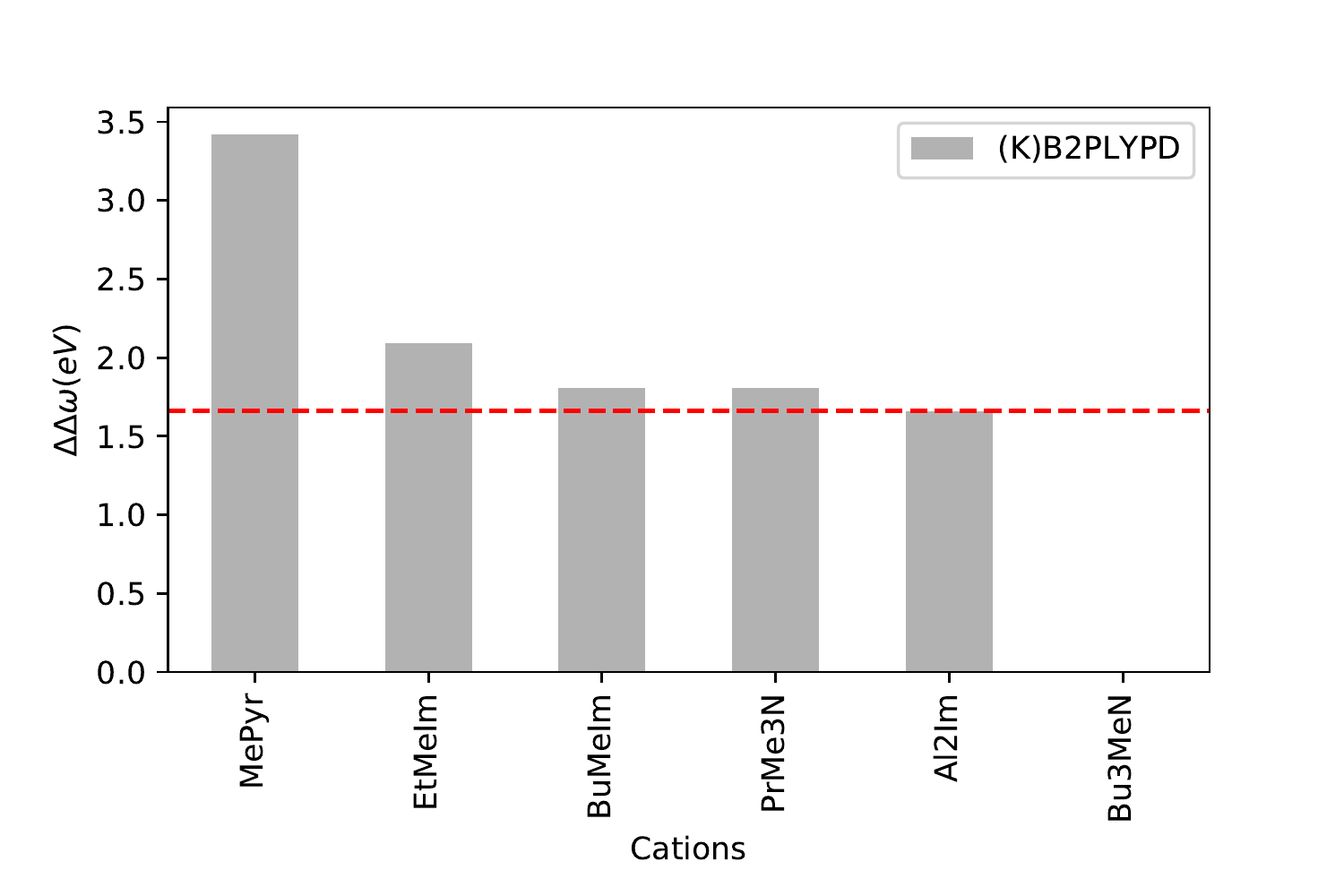} 
\caption{Relative ($\Delta$) net electrophilicity index ($\Delta \omega$) for cations reported in Table \ref{tab:ILsforNP}. The lowest value for $\Delta \omega$ is taken as reference and corresponds to the least reactive cation ($\mathrm{[Bu_3MeN]^+}$). The red dashed line indicates a \textit{hand-made} threshold defined as lowest $\Delta \omega$ amongst the working ILs of Table \ref{tab:ILsforNP}.}
\label{fig:ILsBenchmark}
\end{figure}

\begin{figure}[htp]
\centering
\includegraphics[width=0.5\columnwidth]{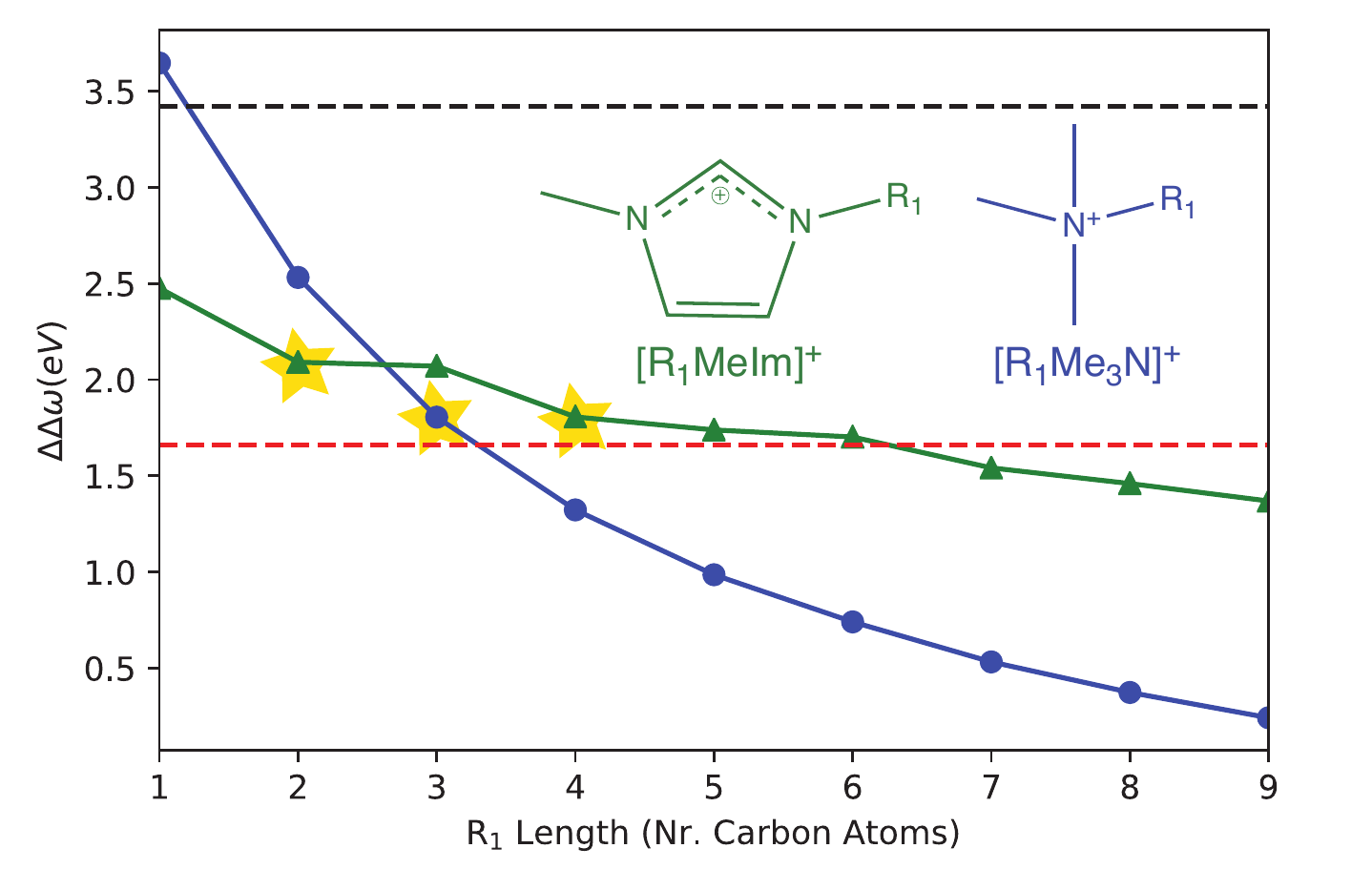} 
\caption{Relative ($\Delta$) net electrophilicity index ($\Delta \omega$) as function of alkyl chain length. The reference is the same of Figure \ref{fig:ILsBenchmark} and corresponds to $\mathrm{[Bu_3MeN]^+}$. In blue $\mathrm{[R_1Me_3N]^+}$ while in green $\mathrm{[R_1MeIm]^+}$ cations. On the $x-$axis, the number of carbon atoms constituting the $\mathrm{R_1}$ chain. The three members of this series ($\mathrm{[BuMeIm]^+}$, $\mathrm{[EtMeIm]^+}$, and $\mathrm{[PrMe_3N]^+}$), already considered in Figure \ref{fig:ILsBenchmark} are marked by yellow, while the empirical effective/non-effective threshold drawn in Figure \ref{fig:ILsBenchmark} is marked with a red-dashed line. The gray-dashed line corresponds to the highest value of $\Delta \omega$ in Figure \ref{fig:ILsBenchmark}. }
\label{fig:ChainLength}
\end{figure}

At first, we performed the calculation of $\Delta \omega$ on the ILs cations already tested in NP production (Figures \ref{fig: cations}-\ref{fig:ILsBenchmark} and Table \ref{tab:ILsforNP}).\\
As shown in Figure \ref{fig:ILsBenchmark}, cations exhibiting higher values of $\Delta \omega$ are the ones that have experimentally shown to be more effective in the NP production and thus radiochemically less stable (Table \ref{tab:ILsforNP}). Vice versa, $\mathrm{[Bu_3MeN]^+}$, the species testified in literature not producing NP, has by far the lowest value of $\Delta \omega$ in this series. Unfortunately, it is cumbersome to find reference to studies of negative cases, namely ILs ineffective in NP production through electron irradiation. Nevertheless, we can comment some of the available (or predicted) trends in literature. Based on the work by Lu et al.,\cite{LU2006140} Tsuda and coworkers\cite{tsuda2012various} predicted that ILs constituted by cations having short alkyl chain, like ethyl groups, enhance crystal growth. We therefore examined the effect of alkyl chain length (Figure \ref{fig:ChainLength}) on the reactivity of alkylmethylimidazolium ($\mathrm{[R_1MeIm]^+}$), and on alkyltrimethylammonium ($\mathrm{[R_1Me_3N]^+}$). We find the expected inverse relation between $\Delta \omega$ and chain length for both types of cations, although more pronounced for $\mathrm{[R_1Me_3N]^+}$  than for  $\mathrm{[R_1MeIm]^+}$, because of the resonance effects. 
Interestingly, we found that in general $\mathrm{[R_1MeIm]^+}$ is more reactive than $\mathrm{[R_1Me_3N]^+}$  with few exceptions corresponding to alkyl chains of only one or two carbons. In particular, we propose the cation $\mathrm{[Me_4N]^+}$ (chain length equal to 1) as good candidate for NP production, because it shows higher $\Delta \omega$ compared to $\mathrm{[MePyr]^+}$  (dashed black line in Figure \ref{fig:ChainLength}), the latter being the most reactive among the cations currently experimentally tested for electron irradiation (Figure \ref{fig: cations}). To the best of our knowledge, $\mathrm{[Me_4N]^+}$ has not been used to produce NPs upon electron irradiation so far, but only e.g. to peptize $\mathrm{Fe_3O_4}$ magnetic NPs through chemical interaction of the cations with the NP surface.\cite{LU2006140, massart1981preparation,andrade2012preparation}

\subsection{Prescreening}

\begin{figure}[htp]
\centering
\includegraphics[width=0.5\columnwidth]{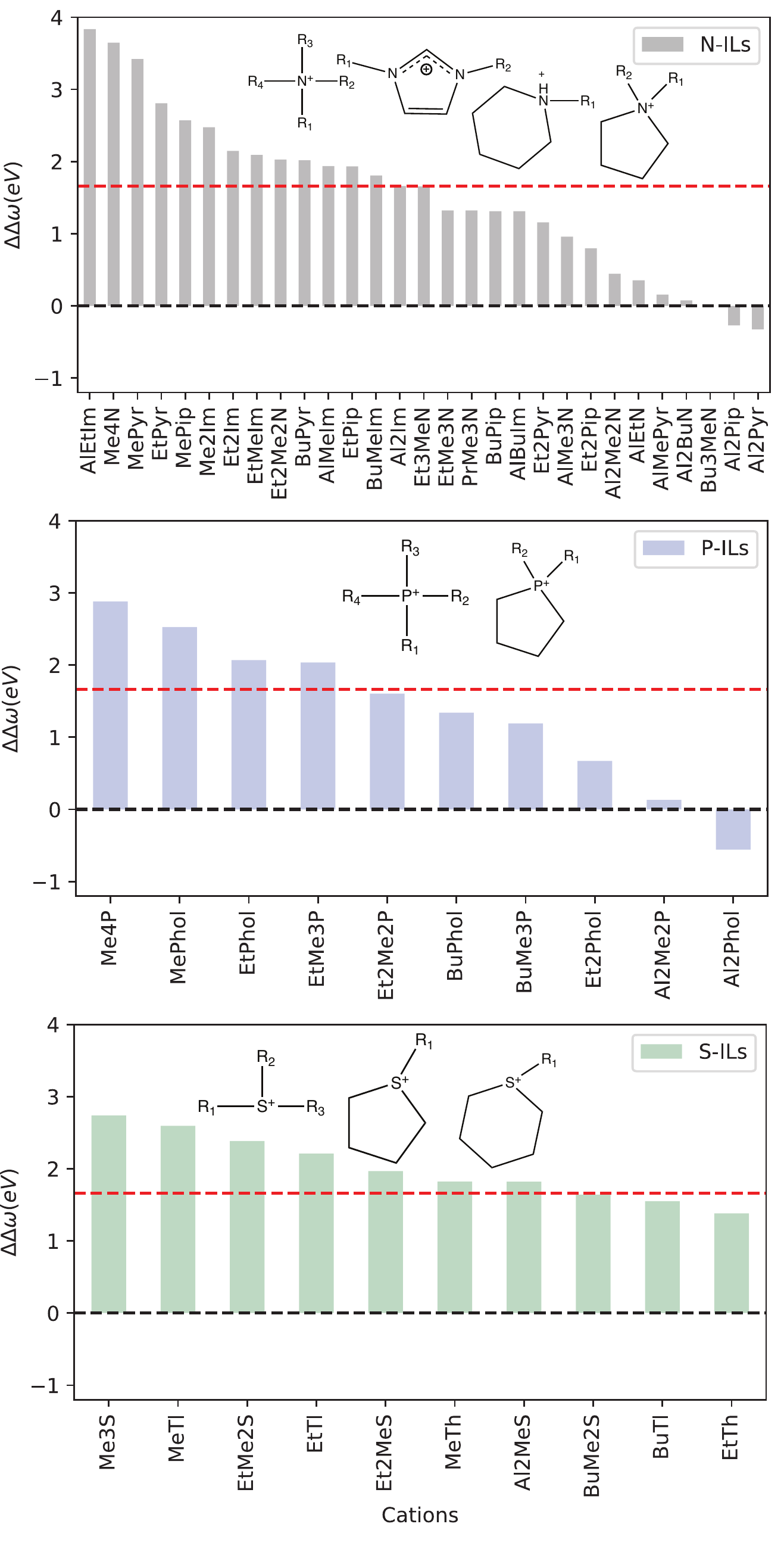} 
\caption{Relative ($\Delta$) net electrophilicity index ($\Delta \omega$) N-based, P-based and S-based cations. The reference (black dashed line) is the same of Figure \ref{fig:ILsBenchmark} and corresponds to $\mathrm{[Bu_3MeN]^+}$. The red dashed line indicates the threshold defined in Figure \ref{fig:ILsBenchmark}. Structures are reported in Figures \ref{fig:Chart_S1}-\ref{fig:Chart_S3} of SI.}
\label{fig:PreScreening}
\end{figure}

The majority of ILs synthesized is based on alkylated amines and imidazolium (N-ILs), and only a few on sulfur (S-ILs) and phosphor (P-ILs) cations. For this reason, there is no indication in literature regarding S-ILs’ and P-ILs’ performances in NP production with electron irradiation techniques. The computational pre-screening in Figure \ref{fig:PreScreening} considers for the first time their possible role in this field by assessing ten possible candidates per category (Figure \ref{fig:Chart_S2}-\ref{fig:Chart_S3} in SI). Twenty N-ILs cations in addition to the ones presented in Figures \ref{fig:ILsBenchmark} are instead considered from N-ILs (\ref{fig:Chart_S1} in SI), yielding a global pre-screening of forty cations. As shown in Figure \ref{fig:ILsBenchmark}, all the cations are characterized by a similar range of $\Delta \omega$ (from about -0.1 to 4.0 eV), and the ones constituting S-ILs and P-ILs seem to be the more stable than the N-based, in line with experimental expectations.\cite{fraser2009phosphonium}

\section{Conclusions}
To conclude, we used the net electrophilicity as a criteria to classify the reactivity of forty cations, constituents of ILs. After successfully benchmarking the protocol against the existing experimental evidence, for each cation category (N-ILs, P-ILs, and S-ILs) we found several good candidates for NP production. Our prediction should be tested experimentally relatively straightforwardly. The computational protocol can be easily extended to thousands of species, opening new horizons e.g. for NP synthesis observation by in situ TEM in ILs, and in general for NP synthesis in ILs including all their possible applications, from catalysis to electrochemistry.

\newpage

\newpage

\makeatletter

\section{Supporting Information}
\setcounter{section}{1}
\setcounter{table}{0}
\setcounter{equation}{0}
\setcounter{figure}{0}

\renewcommand \thesection{S\@arabic\c@section}
\renewcommand\thetable{S\@arabic\c@table}
\renewcommand \thefigure{S\@arabic\c@figure}
\renewcommand \theequation{S\@arabic\c@equation}
\makeatother

\subsection{Theory: Conceptual DFT}
\label{section: theory}
\subsubsection{Net Electrophilicity}
In the following, we report the different definitions\cite{chattaraj2009net} for $\Delta \omega$, $\omega^+$ and $\omega^-$. 
\begin{equation}
\label{eq:1SI}
\Delta \omega = \omega^+ +  -(-\omega^-)=\omega^+ + \omega^-,
\end{equation}
\begin{equation}
\label{eq:2SI}
\Delta \omega = \omega^+ +  \left(\frac{1}{\omega^-}\right).
\end{equation}
\textbf{Def 1} for $\omega^+$ and $\omega^-$ :
\begin{equation}
\label{eq:3SI}
\omega^-=\frac{(3IP+EA)^2}{16(IP-EA)},
\end{equation}
\begin{equation}
\label{eq:4SI}
\omega^+=\frac{(IP+3EA)^2}{16(IP-EA)}.
\end{equation}
\textbf{Def 2} for $\omega^+$ and $\omega^-$ :
\begin{equation}
\label{eq:5SI}
\omega^-=\frac{(IP)^2}{2(IP-EA)},
\end{equation}
\begin{equation}
\label{eq:6SI}
\omega^-=\frac{(EA)^2}{2(IP-EA)}.
\end{equation}
IP, and EA are the ionization potential and electron affinity, whose calculation is discussed in \ref{subsection: IP&EA}. 

\subsubsection{Ionization Potentials and Electron Affinities}
\label{subsection: IP&EA}
The accuracy of the calculation of IP and EA is expected to vary depending both on the level of theory and on the basis set.
For Hartree Fock calculations, IP and EA can be calculated as the negative of the energies of the HOMO and LUMO respectively, using Koopmans’ theorem.\cite{koopmans1934zuordnung} 
For Density Functional Theory (DFT) methods, the HOMO and LUMO orbitals are well known not to accurately reproduce IPs without corrections.\cite{cooper2012ab} This limitation can be overcome calculating the IP and EA from the absolute energies between the N, N+1, and N-1 electrons system:
\begin{equation}
\label{eq:7SI}
IP=E(N-1)-E(N),
\end{equation}
\begin{equation}
\label{eq:8SI}
EA=E(N)-E(N+1).
\end{equation}

\subsection{Methods}
Calculations were performed with GAMESS code.\cite{gordon2005advances} At first, geometry optimizations are carried out at B3LYP/6-311+G(d,p).\cite{10.1063/1.464913} Although DFT B3LYP functional has been successfully used in a wide range of applications, after the geometry optimizations, we chose to carry out more accurate energy calculations at the B2PLYPD level,\cite{grimme2006semiempirical} proven to outperform the other DFT and some beyond-DFT methods (MP2) especially in the estimation of EA.\cite{cooper2012ab} 

\subsubsection{Simulated Systems}
\begin{table}[htp]
\centering
\caption{Abbreviations used for the cations}
\label{tab:ILsabbreviations}
\begin{tabular}{cccc}
\toprule
Abbreviation &  & Abbreviation &  \\
\midrule
Al & ally & Phol & pholanium \\
Bu & buthyl & Pr & propyl \\
Et & ethyl & Pyr & pyrrolidinium \\
Im & imidazolium	 & S & sulfonium \\
Me & methyl & ThPyr &	thiopyrylium \\
N & ammonium & Tl & thiolanium \\
P & phosphonium & \\
\bottomrule
\end{tabular}
\end{table}

\begin{figure}[htbp]
    \centering
    \includegraphics[width=0.5\columnwidth]{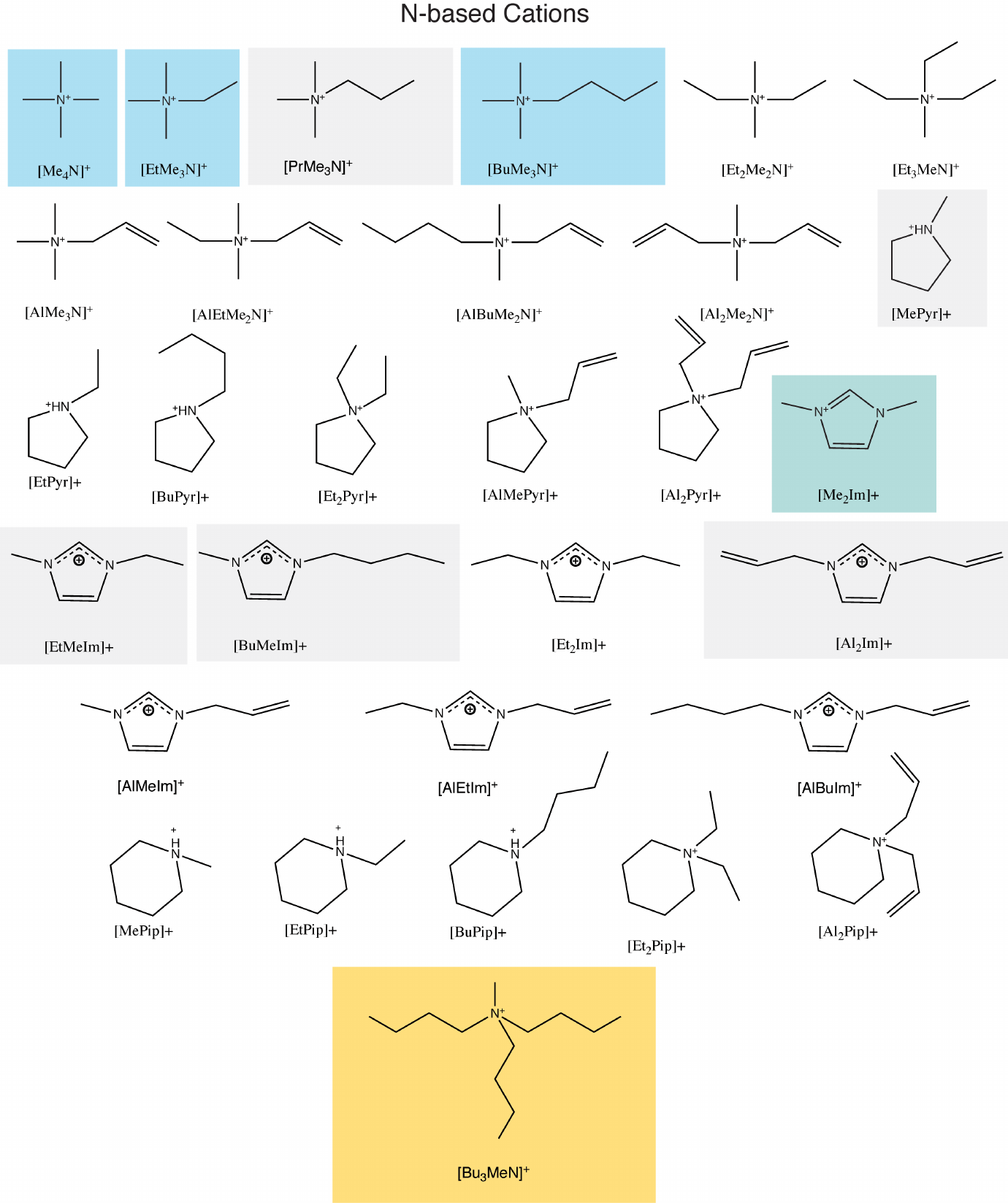}
    \caption{Schematic of the N-ILs cations considered. Gray-highlighted structures are reported in the benchmark cal-culation of Figure 1. Orange-highlighted cation is the reference ($\Delta \omega=0$) for Figures \ref{fig:ILsBenchmark}-\ref{fig:PreScreening} of the manuscript. Green and blue-highlighted cations are reported in Figure \ref{fig: cations} of the manuscript.}
    \label{fig:Chart_S1}
\end{figure}

\begin{figure}[htbp]
    \centering
    \includegraphics[width=0.5\columnwidth]{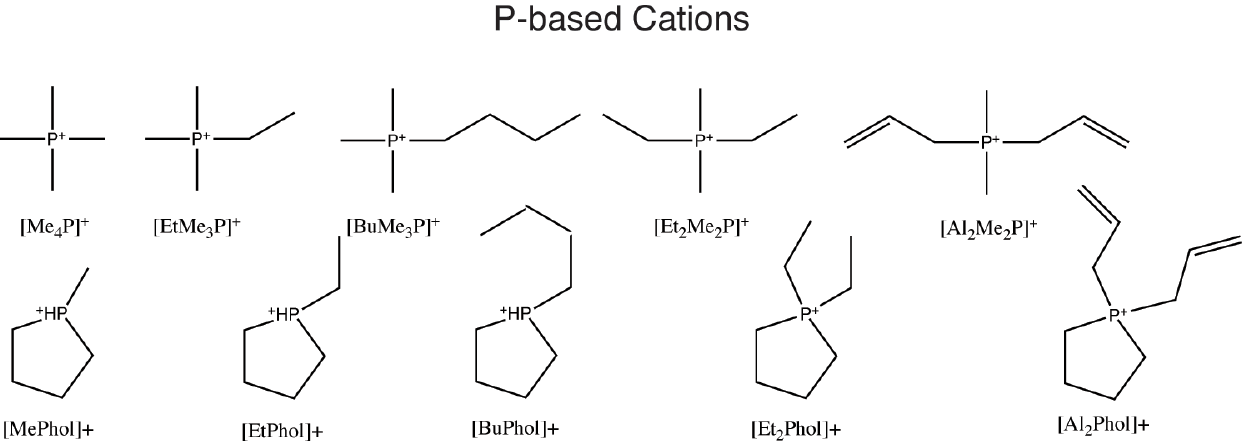}
    \caption{Schematic of the P-ILs cations considered. The corresponding $\Delta \omega$ is reported in Figure \ref{fig:PreScreening}.}
    \label{fig:Chart_S2}
\end{figure}

\begin{figure}[htbp]
    \centering
    \includegraphics[width=0.5\columnwidth]{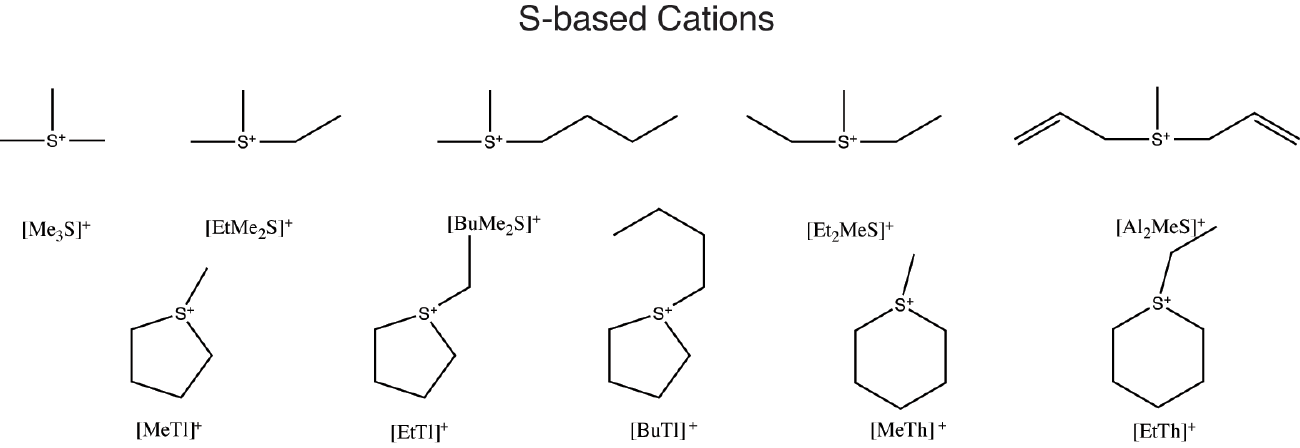}
    \caption{Schematic of the S-ILs cations considered. The corresponding $\Delta \omega$ is reported in Figure \ref{fig:PreScreening}.}
    \label{fig:Chart_S3}
\end{figure}

\subsection{Preliminary Tests}
\label{section: test definitions}
\subsubsection{On the Different Definitions of Net Electrophilicy}
\begin{figure}[ht]
    \centering
    \includegraphics[width=0.5\columnwidth]{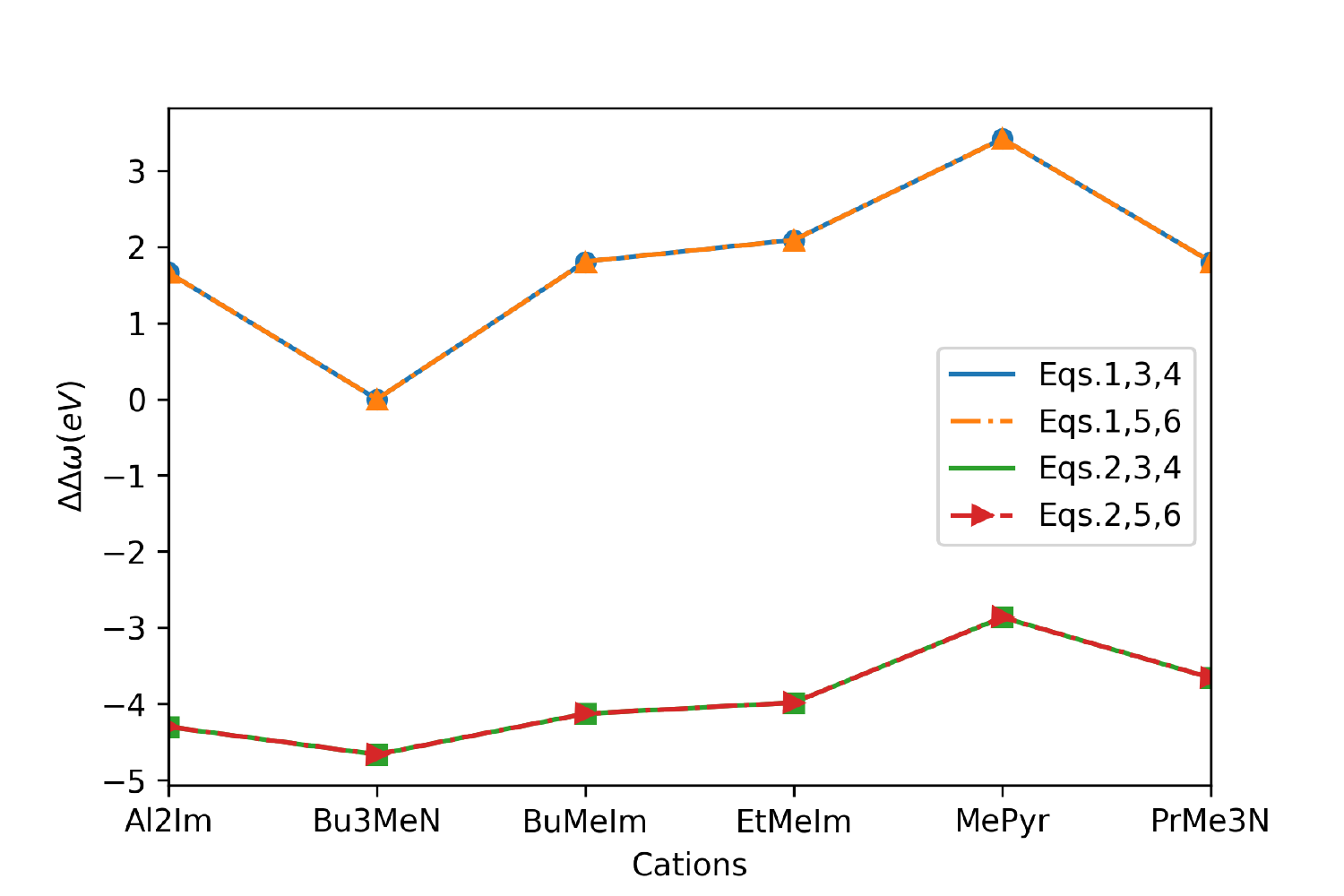}
    \caption{Differences in Net Electrophilicity (eV). The lowest $\Delta \omega$ calculated using Eqs. \ref{eq:1SI}, \ref{eq:3SI}-\ref{eq:4SI} has been taken as reference. IP and EA have been calculated using Koopmans’ theorem using HOMO-LUMO energies calculated at B2PLYPD level. }
    \label{fig:Figure_S1}
\end{figure}

On the six cations reported in the main paper, we tested the validity of the definitions reported in Eqs. \ref{eq:1SI}-\ref{eq:6SI}. As shown in Figure \ref{fig:Figure_S1}, although Eqs. \ref{eq:1SI} and \ref{eq:2SI} give rise to different values of $\Delta \Delta \omega$, a similar trend is recognized. The use of Eqs. \ref{eq:5SI},\ref{eq:6SI} in place of Eqs. \ref{eq:3SI},\ref{eq:4SI} does not affect the results.

\subsubsection{Net Electrophilicy at Different Levels of Theory}
\begin{figure}[htpb]
    \centering
    \includegraphics[width=0.5\columnwidth]{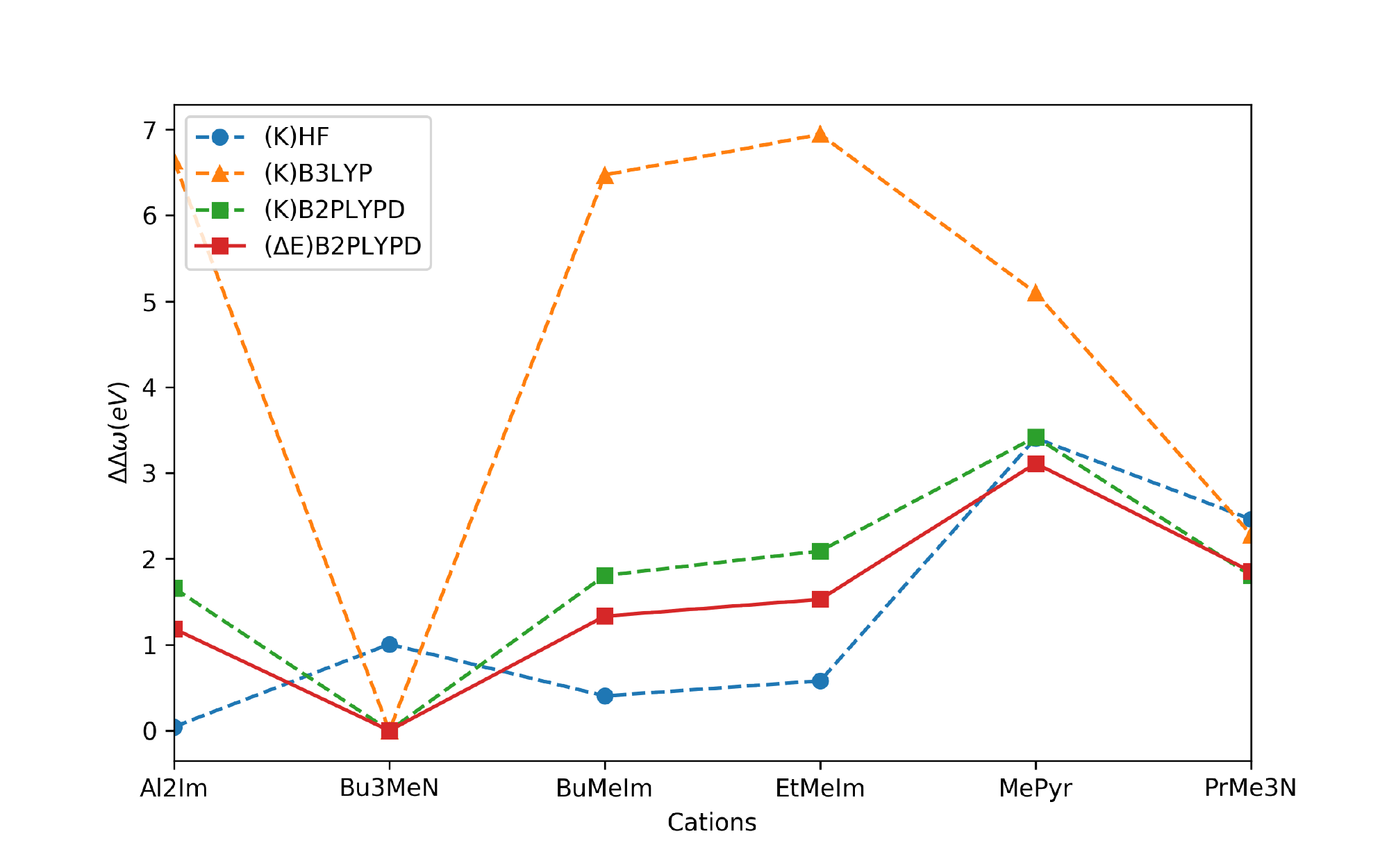}
    \caption{Differences in Net Electrophilicity (eV). The lowest $\Delta \omega$ value for each level of theory has been taken as reference. Legend: (K) indicates that IP and EA have been calculated using Koopmans’ theorem. ($\Delta E$) indicates that they have been evaluated accordingly to Eqs. \ref{eq:6SI}-\ref{eq:7SI}.}
    \label{fig:levelstheory}
\end{figure}

For the six cations reported in Table \ref{tab:ILsforNP} of the manuscript, we tested the estimation of the net electrophilicity at different levels of theory, namely HF, DFT/B3LYP,\cite{10.1063/1.464913} and DFT/B2PLYPD.\cite{grimme2006semiempirical} The geometries were optimized with DFT/B3LYP functional. The calculation of $\Delta \omega$ was performed estimating IP and EA using Koopmans’ theorem and through the differences of absolute energies (B3LYP and B2LYP). The results are reported in Figure \ref{fig:levelstheory}.
In our opinion, the use of Koopmans’ theorem is not recommended if approximating DFT exchange-correlation functional with B3LYP, because it predicts $\Delta \Delta \omega$ values in a range that significantly differs from the ones of the values resulting from HF and DFT/B2PLYPD calculations. 
The calculation of IP and EA using Koopmans’ theorem at DFT/B2PLYPD level gives similar results to the dif-ferences in absolute energies (cfr. the dashed green and the solid red lines), hence it has been used for the calcula-tion of the net electrophilicity of the radical species. 
HF and DFT/ B2PLYPD show similar trends, with the exception of one cation, $\mathrm{[Bu3MeN]^+}$ and $\mathrm{[Al2Im]^+}$.
We analyzed this result comparing HOMO and LUMO orbitals. 

\subsubsection{The Effect of the Implicit Solvent}
\begin{figure}[htpb]
    \centering
    \includegraphics[width=0.5\columnwidth]{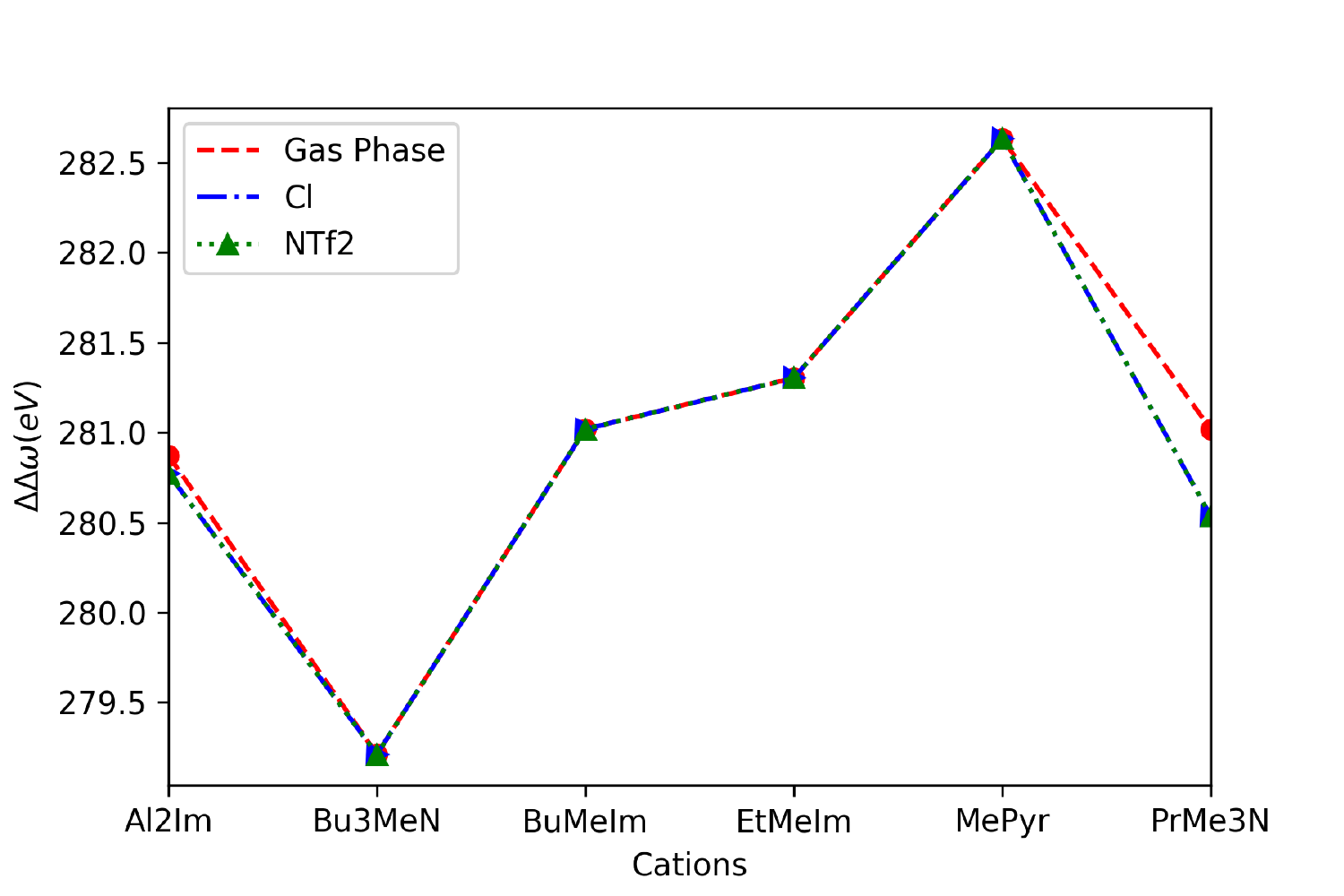}
    \caption{The lowest $\Delta \omega$ value in gas phase (corresponding to $\mathrm{[Bu3MeN]^+}$) was taken as reference. Parameters describing the SMD-GIL are reported in Table \ref{tab:implicit_solvent}.}
    \label{fig: implicitsolvent}
\end{figure}

In the following, we comment the effect of the SMD-GIL implicit solvent developed by Bernales and coworkers\cite{bernales2012quantum} on the estimation of $\Delta \omega$. We modified the parameters for the aromaticity ($\Phi$) and halogenicity ($\Psi$), accordingly to the IL, keeping all the other parameters unchanged.
For the calculation of $\Phi$ and $\Psi$, we considered two different anions, to test if the trend for $\Delta \omega$ is maintained, taking as reference the gas phase case. 
We conducted this test on the six cations reported in Table \ref{tab:implicit_solvent}, performing energy calculations at B2PLYPD level on the geometries optimized with the B3LYP functional in presence of the SMD-GIL solvent. As shown in Figure \ref{fig: implicitsolvent}, no significant differences from the calculation in gas phase are observed. EA and IP were calculated again using Koopmans’ theorem.

\begin{table}[htbp]
  \centering
  \caption{Parameters cation/anion dependent for the SMD-GIL implicit solvent.\\
Additional parameters: $\sum \alpha_2^H=0.229$, $\sum \beta_2^H=0.265$, $\varepsilon=11.50$, $n=1.43$, $\gamma=61.24$}
    \begin{tabular}{p{15.75em}cccc}
    \toprule
    \multicolumn{1}{l}{Anion represented by the SMD-GIL $\rightarrow$} & \multicolumn{2}{c}{$\mathrm{[NTf2]^-}$} & \multicolumn{2}{c}{$\mathrm{Cl^-}$} \\
    Cation $\downarrow$ & \multicolumn{1}{p{5em}}{} & \multicolumn{1}{p{5em}}{} & \multicolumn{1}{p{5em}}{} & \multicolumn{1}{p{5em}}{} \\
    \midrule
    $\mathrm{[Al2Im^]+}$ & 0.12  & 0.23  & 0.25  & 0.08 \\
    $\mathrm{[Bu3MeN]^+}$ & 0     & 0.21  & 0     & 0.07 \\
    $\mathrm{[BuMeIm]^+}$ & 0.12  & 0.24  & 0.27  & 0.09 \\
    $\mathrm{[EtMeIm]^+}$ & 0.13  & 0.26  & 0.33  & 0.11 \\
    $\mathrm{[MePyr]^+}$ & 0.19  & 0.29  & 0.57  & 0.14 \\
    $\mathrm{[PrMe3N]^+}$ & 0     & 0.27  & 0     & 0.13 \\
    \bottomrule
    \end{tabular}%
  \label{tab:implicit_solvent}%
\end{table}%

\newpage
\bibliographystyle{elsarticle-num}
\bibliography{references}  
\end{document}